\documentclass{INTERSPEECH2023}
\usepackage[hang,flushmargin]{footmisc}
\usepackage{pifont}
\usepackage{makecell}
\newcommand{\tabincell}[2]{\begin{tabular}
{@{}#1@{}}#2\end{tabular}}
\newcommand{\xmark}{\ding{55}}

\usepackage{multirow}
\usepackage{cite}
\usepackage{amsmath}
\usepackage[font=small,labelfont=bf]{caption}

\interspeechcameraready

\title{Factorised Speaker-environment Adaptive Training\\ of Conformer Speech Recognition Systems}
\name{Jiajun Deng$^1$, Guinan Li$^1$, Xurong Xie$^2$, Zengrui Jin$^1$, Mingyu Cui$^1$,\\ Tianzi Wang$^1$, Shujie Hu$^1$, Mengzhe Geng$^1$, Xunying Liu$^1$}

\address{
  $^1$The Chinese University of Hong Kong, Hong Kong SAR, China\\
  $^2$Institute of Software, Chinese Academy of Sciences, Beijing, China}
\email{\{jjdeng,gnli,zrjin,mycui,twang,sjhu,mzgeng,xyliu\}@se.cuhk.edu.hk, xurong@iscas.ac.cn}
\begin{document}
\bstctlcite{IEEEexample:BSTcontrol}
\maketitle

\begin{abstract}
Rich sources of variability in natural speech present significant challenges to current data intensive speech recognition technologies. To model both speaker and environment level diversity, this paper proposes a novel Bayesian factorised speaker-environment adaptive training and test time adaptation approach for Conformer ASR models. Speaker and environment level characteristics are separately modeled using compact hidden output transforms, which are then linearly or hierarchically combined to represent any speaker-environment combination. Bayesian learning is further utilized to model the adaptation parameter uncertainty. Experiments on the 300-hr WHAM noise corrupted Switchboard data suggest that factorised adaptation consistently outperforms the baseline and speaker label only adapted Conformers by up to 3.1\% absolute (10.4\% relative) word error rate reductions. Further analysis shows the proposed method offers potential for rapid adaption to unseen speaker-environment conditions. 

\end{abstract}
\section{Introduction}
The majority of end-to-end (E2E) automatic speech recognition (ASR) systems\cite{li2022recent}, including those based on state-of-the-art Conformer models\cite{guo2021recent}, are usually trained and evaluated on found speech data collected from a wide range of real-world scenarios. Such naturalistic speech data is generally highly non-homogeneous. Rich sources of variability are brought by multiple acoustic factors\cite{benzeghiba2006impact}, for example, speaker characteristics, background noise and recording channel conditions. The resulting high degree of speech heterogeneity presents significant challenges to current data intensive speech recognition technologies in multiple stages. These include both the construction of speaker and environment independent ASR systems, and their fine-grained adaptation to individual users’ voice recorded in diverse acoustic environments. 

Prior researches in this direction to date have been largely spearheaded into two separate areas with their respective focuses on either speaker adaptation \cite{bell2020adaptation}, or speech enhancement and environment compensation only\cite{li2014overview}. In the first area, auxiliary speaker-aware features that are based on i-vector\cite{karafiat2011ivector,tuske2021limit,zeineldeen2022improving}, x-vector\cite{zeineldeen2022improving,baskar2022speaker}, feature-space maximum likelihood linear regression (f-MLLR)\cite{gales1998maximum,baskar2022speaker}, or extracted from speaker-aware modules\cite{delcroix2018auxiliary,zhao2020speech,sari2020unsupervised} are incorporated into various ASR models. Model-based speaker adaptation methods estimate speaker-dependent (SD) parameters, which are implemented as either internal DNN components\cite{ochiai2018speaker,huang2021rapid}, or additional parameters such as learning hidden unit contributions (LHUC)\cite{Swietojanski2016LearningHU,Xie2021BayesianLF,deng2023confidence}, using the target speaker data during speaker adaptive training and test time adaptation\cite{anastasakos1996compact,ochiai2014speaker}. In the second area, single-channel based environment compensation \cite{moreno1996vector,stouten2006model,yu2008minimum,yoshioka2015environmentally,ravanelli2020multi} or multi-channel based speech enhancement front-ends\cite{seltzer2004likelihood,anguera2007acoustic,xu2019joint,heymann2019joint,yu2021audio,zhang2021end} are separately constructed and optionally further integrated with the recognition back-end. Back-end model adaptation methods that aim to compensate for the modelling mismatch against the unseen target environment have also been studied\cite{seltzer2010acoustic,chen2014initial}.

A simple approach to handle the multifaceted data heterogeneity in natural speech is to separately model each user’s voice recorded in diverse environments as different speakers. However, this fails to account for the homogeneity over speaker-level characteristics, leading to fragmentation of data and poor generalization to unseen speaker-environment combinations. 

An alternative and more general solution to such a problem is to structurally represent different factors of variability in ASR systems\cite{gales2001acoustic,wang2013explicit,seltzer2012factored,fainberg2017factorised,kitza2019cumulative}. For example, during the adaptive training stage \cite{anastasakos1996compact,gales2001adaptive}, speaker and environment characteristics are “factored out” into their respective separately designed modelling components (e.g., vector Taylor series \cite{acero2000hmm}, MLLR or CMLLR \cite{leggetter1995maximum,gales1998maximum}, or LHUC \cite{Swietojanski2016LearningHU} transforms), thus the backbone ASR model can focus more on learning speaker and environment invariant speech representations and their mapping to spoken contents. During the test time adaptation stage, these sources of variabilities can be flexibly “factored in” to model any seen or unseen speaker-environment combination. Prior researches in this direction were mainly conducted for conventional GMM-HMM \cite{gales2001acoustic,seltzer2011separating,wang2012speaker,wang2013explicit,seltzer2012factored} and hybrid DNN-HMM \cite{fainberg2017factorised,kitza2019cumulative} ASR systems. In contrast, existing researches on E2E ASR systems represented by Conformer largely focus on modelling only one source of variability, for example, speaker characteristics\cite{tuske2021limit,zeineldeen2022conformer,zeineldeen2022improving,deng2023confidence}, or environmental mismatch\cite{chang2020end,zhang2021end,kumar2022end,sukhadia2023domain}. 

To this end, a novel factorised speaker-environment adaptive training approach is proposed in this paper to facilitate both adaptive training and test time unsupervised adaptation of E2E Conformer models. Speaker and environment level characteristics are separately modelled using compact LHUC \cite{Swietojanski2016LearningHU} or hidden unit bias (HUB) \cite{abdel2013fast,deng2023confidence} transformations. These are linearly or hierarchically combined to represent any speaker-environment combination, observed in the training data or otherwise. Bayesian estimation of the speaker or environment factor specific transforms is also utilized to mitigate the risk of overfitting during test time unsupervised adaptation to the limited speaker or environment data. The acquired speaker and environment homogeneity can be exploited for rapid adaptation to the unseen speaker-environment combination. For example, speaker specific transforms estimated in one environment can be cached and reused in another environment. The main contributions of the paper are summarized below: 

{\bf 1)} To the best of our knowledge, this paper presents the first work to investigate the model-based factorised adaptation for E2E Conformer models by structurally representing speaker and environment factors of variability. In contrast, prior researches on E2E ASR systems largely focus on modelling only one source of variability\cite{zhao2020speech,tuske2021limit,huang2021rapid,zeineldeen2022conformer,zeineldeen2022improving,deng2023confidence,chang2020end,zhang2021end,kumar2022end,sukhadia2023domain}. 

{\bf 2)} The efficacy of the proposed Bayesian factorised adaptation approaches is consistently demonstrated on the 300-hr WHAM noise corrupted Switchboard task. Experimental results suggest that our approach consistently outperforms the un-adapted baseline and speaker label only adapted Conformer systems by up to 3.1\%, 2.7\% and 2.9\% absolute (10.4\%, 8.1\%, and 8.2\% relative) word error rate (WER) reductions on the noise corrupted Hub5’00, RT02, and RT03 test sets respectively, before and after external language model rescoring is applied.

{\bf 3)} Further analysis shows that the proposed method offers the potential for rapid adaptation to the unseen speaker-environment combination by flexibly “factoring in” the already estimated speaker and environment specific transforms. Furthermore, their generic nature and the implementation details described in this paper allow their further application to handle two or more sources of variability in other E2E ASR tasks.

\section{Conformer E2E ASR System}
\begin{figure}
    \centering
    \setlength{\abovecaptionskip}{0pt plus 0pt minus 0pt}
    \includegraphics[scale=0.45]{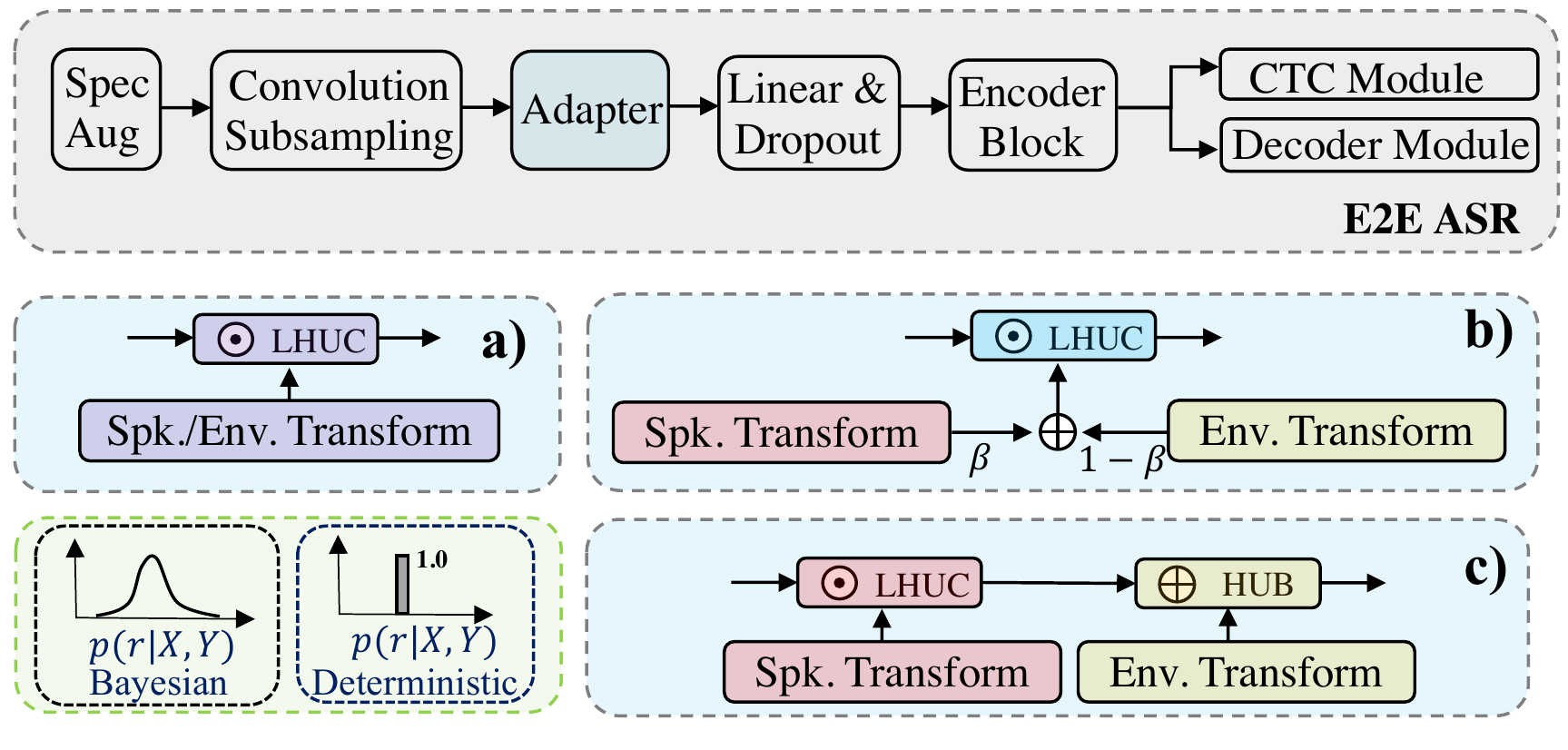}
    \caption{Examples of Conformer E2E ASR models (grey box, top), together with three model-based adaptation methods: \textbf{a)} Conformer speaker adaptation using LHUC transforms; \textbf{b)} linear (superposition) factorised adaptation; and \textbf{c)} cascaded factorised adaptation. Bayesian and deterministic estimations of adaptation parameters are shown in the green box (bottom left).} 
    \label{fig:adaptation}
    \vspace{-0.4cm}
\end{figure}

The Conformer \cite{guo2021recent} ASR model consists of an encoder module and a decoder module, which are both based on multi-blocked stacked architectures. The  encoder module comprises a convolutional subsampling module, a linear layer with dropout operation, and stacked encoder blocks. Layer normalization and residual connections are performed on all encoder blocks. More details of Conformer components can be found in \cite{gulati2020conformer}. Fig.~\ref{fig:adaptation} shows an example of Conformer E2E ASR system.

For training the Conformer model, the following multi-task criterion interpolation between connectionist temporal classification (CTC) and attention error cost is adopted \cite{watanabe2017hybrid}.
\begin{equation}
{\cal{L}}=(1-\lambda) {{\cal{L}}_{att}} + \lambda{{\cal{L}}_{ctc}},
\label{enq:loss_combine}
\end{equation}
where $\lambda \in [0, 1]$ is a tunable hyper-parameter and empirically set as 0.2 for training and 0.3 for recognition in this paper.

\section{Conformer Speaker Adaptation}
The key idea of LHUC adaptation \cite{Swietojanski2016LearningHU,deng2023confidence} is to use the SD scaling vector to modify the amplitudes of activation outputs. Let $\bm{r}^{l,s}$ denote the SD parameters for speaker $s$ in the $l$-th hidden layer, the speaker adapted hidden outputs can be given by
\begin{align}
{\bm{h}}^{l,s}= {\bm{h}}^{l} \odot \xi(\bm{r}^{l,s}), \label{lhuc_spk}
\end{align}
where $\bm{h}^{l}$ is the hidden activation outputs in the $l$-th hidden layer, $\odot$ is the Hadamard product operation, and $\xi(\cdot)$ is the element-wise 2$\times$ Sigmoid($\cdot$) function. 

Alternatively the SD transform that is added to the hidden output as a bias vector \cite{abdel2013fast,deng2023confidence} leads to the HUB adaptation. Let $\zeta({\bm r}^{l,s})$ denote the bias vector for speaker $s$ in the $l$-th hidden layer. The speaker HUB adapted hidden outputs are given as
\begin{align}
\setlength{\abovedisplayskip}{1pt plus 1pt minus 0pt}
\setlength{\belowdisplayskip}{1pt plus 1pt minus 0pt}
{\bm{h}}^{l,s}= {\bm{h}}^{l} + \zeta(\bm{r}^{l,s}), \label{hub_spk}
\end{align}
where $\zeta(\cdot)$ is the identity activation function.

LHUC and HUB can be further used for environment adaptation by learning an environment specific LHUC or HUB transform, and a single joint speaker-environment transform to model a particular combination of these two factors.  

\vspace{-0.2cm}
\section{Factorised Conformer Speaker-environment Adaptation}
\vspace{-0.05cm}
\subsection{Linear Factorised Adaptation}
\vspace{-0.05cm}
Linear factorised adaptation (LFA) models the two acoustic factors using a linear interpolation between a speaker-dependent (SD) transform and an environment-dependent (ED) transform, as is shown in Fig. \ref{fig:adaptation}(b). Let $\bm{n}^{l,e}$ denote the ED parameters for environment $e$ in the $l$-th hidden layer. The factorised adapted hidden outputs for speaker $s$ in  environment $e$ can be derived by 
\begin{align}
{\bm{h}}^{l,s,e}= {\bm{h}}^{l} \odot (\beta \xi(\bm{r}^{l,s}) + (1-\beta)\xi(\bm{n}^{l,e})),
\end{align}
where $\beta \in [0, 1]$ is a hyper-parameter that balances the weighting between the speaker and environment factors. For example, $\beta=1$ and $\beta=0$ lead to the LHUC speaker only adaptation and environment only adaptation, respectively. 
\vspace{-0.2cm}
\subsection{Cascaded Factorised Adaptation}
In the cascaded factorised adaptation (CFA), the SD transform and ED transform, which serve as either an LHUC scaling vector or a HUB bias vector, are cascaded into the Conformer hidden layers. This leads to the following four cases: 1) both transforms are LHUC scaling vectors; 2) both transforms are HUB bias vectors; 3) the SD transform is an LHUC scaling vector while the ED transform is a HUB bias vector; and 4) the SD transform is a HUB bias vector while the ED transform is an LHUC scaling vector. For example, without loss of generality, assuming that both transforms are applied at the same layer, the above third case is shown in Fig. \ref{fig:adaptation}(c) and the factorised adapted hidden outputs can be given by  
\begin{align}
{\bm{h}}^{l,s,e}= {\bm{h}}^{l} \odot \xi(\bm{r}^{l,s}) + \zeta(\bm{n}^{l,e}).
\end{align}

\vspace{-0.3cm}
\subsection{Estimation of Factorised Adaptation Parameters}
Let ${\cal{D}}^{s,e}=\{\bm{X}^{s,e},\bm{Y}^{s,e}\}$ denote the data set for speaker $s$ in the environment $e$, where $\bm{X}^{s,e}$ and $\bm{Y}^{s,e}$ are the acoustic features and the corresponding supervision token sequences, respectively. During unsupervised test time adaptation, the supervision $\bm{Y}^{s,e}$ of unseen test data need to be generated by initially decoding the corresponding utterances using an un-adapted baseline Conformer model, before serving as the target token labels in the subsequent adaptation. The SD and ED parameters can be estimated by using the loss in Eqn.~(\ref{enq:loss_combine}), 
\begin{align}
\setlength{\abovedisplayskip}{0pt plus 1pt minus 2pt}
\setlength{\belowdisplayskip}{0pt plus 1pt minus 2pt}
\{\hat{{\bm r}}^s, \hat{{\bm n}}^e\} = \mathop{\arg\min}\limits_{\{{\bm r}^s, {\bm n}^e\}}\{{\cal L}(\bar{\cal D}^{s,e};{\bm r}^s, {\bm n}^e)\},
\end{align}
where $\bar{\cal D}^{s,e}$ is the union of all speaker's adaptation data in a given environment $e$, $\mathop{\cup}_{i\in{\cal S}}{\cal{D}}^{i,e}$, and all environment's adaptation data associated with a speaker $s$, $\mathop{\cup}_{i\in{\cal E}}{\cal{D}}^{s,i}$. 

During adaptive training, the SD and ED parameters associated with the training data are jointly optimized with the "canonical" model parameters $\bm{\Theta}$ that are independent of speaker or environment characteristics. This is given as  
\begin{align}
\setlength{\abovedisplayskip}{0pt plus 1pt minus 2pt}
\setlength{\belowdisplayskip}{0pt plus 1pt minus 2pt}
\{\hat{{\bm{\Theta}}}, \hat{\bm{\theta}}_{S}, \hat{\bm{\theta}}_{E}\} = \mathop{\arg\min}_{\{{{\bm{\Theta}}}, {\bm{\theta}}_{S}, {\bm{\theta}}_{E}\} }\sum_{s\in{\cal S}}\sum_{e\in{\cal E}} {\cal L}({\cal D}^{s,e}; {{\bm{\Theta}}}, {\bm{\theta}}_{S}, {\bm{\theta}}_{E}), 
\end{align}
where $\bm{\theta}_{S}=\{{\bm r}^{s}\}_{s\in{\cal S}}$ and $\bm{\theta}_{E}=\{{\bm n}^{e}\}_{e\in{\cal E}}$ are the SD and ED parameter sets associated with training data, respectively. 

\vspace{-0.2cm}
\subsection{Bayesian Learning of Factorised Adaptation}
Bayesian learning\cite{mackay1992practical} is adopted to model adaptation parameter uncertainty. Given limited adaptation data $\bar{{\cal D}}^{s,e}$, the Bayesian predictive distribution for a test utterance $\tilde{\bm{X}}^{s,e}$ is $\iint{p(\tilde{\bm{Y}}^{s,e}|\tilde{\bm{X}}^{s,e},{\bm r}^s,{\bm n}^e)p({\bm r}^s,{\bm n}^e|\bar{{\cal D}}^{s,e})d{\bm r}^s}d{\bm n}^e$,
where ${\tilde{\bm Y}}^{s,e}$ is the predicted token sequence and $p({\bm r}^s,{\bm n}^e|\bar{{\cal D}}^{s,e})$ is the joint posterior distribution of the SD and ED parameters learned from the adaptation data. Using variational inference, a variational distribution $q(\bm{r}^s,\bm{n}^e)$ is used to approximate the joint posterior distribution $p({\bm r}^s,{\bm n}^e|\bar{{\cal D}}^{s,e})$, and inferred by optimizing the hybrid attention plus CTC loss marginalization ${\cal L}(\bar{\cal {D}}^{s,e})=(\lambda-1)\log\iint p_a(\bm{r}^s, \bm{n}^e|\bar{\cal {D}}^{s,e})d\bm{r}^sd\bm{n}^e-\lambda\log\iint p_c(\bm{r}^s, \bm{n}^e|\bar{\cal {D}}^{s,e})d\bm{r}^sd\bm{n}^e$ over the uncertain parameters, ${\bm r}^s,{\bm n}^e$. The variational bound is given by
\begin{align}
\setlength{\abovedisplayskip}{2pt plus 1pt minus 1pt}
\setlength{\belowdisplayskip}{2pt plus 1pt minus 1pt}
&{\cal L}(\bar{\cal {D}}^{s,e})\leq \iint q(\bm{r}^s,\bm{n}^e)\{(\lambda-1)\log p_a(\bar{\cal {D}}^{s,e}|\bm{r}^s, \bm{n}^e) - \nonumber\\
&\lambda\log p_c(\bar{\cal {D}}^{s,e}|\bm{r}^s, \bm{n}^e)\}d\bm{r}^sd\bm{n}^e + \text {KL}(q(\bm{r}^{s}, \bm{n}^e)||p({\bm r}^s,{\bm n}^e)) \nonumber\\
&\triangleq {\cal L}_{int}(\bar{\cal {D}}^{s,e};\bm{r}^s, \bm{n}^e) + {\cal L}_{KL},
\end{align}
where $p_a$ and $p_c$ are the attention and CTC based sequence probabilities respectively, $p({\bm r}^s,{\bm n}^e)$ is the joint prior distribution of the SD and ED parameters. KL$(\cdot)$ is the KL divergence. Since the SD and ED latent variables $\{\bm{r}^s,\bm{n}^e\}$ are independent of each other, the joint variational and prior distributions can be modeled independently. The structured variational distributions $\{q(\bm{r}^s)={\cal N}(\bm{\mu}_r^{s},\bm{\sigma}_r^{s}),q(\bm{n}^e)={\cal N}(\bm{\mu}_n^{e},\bm{\sigma}_n^{e})\}$ and the prior distributions $\{p(\bm{r}^s)={\cal N}(\bar{\bm{\mu}}_r,\bar{\bm{\sigma}}_r),p(\bm{n}^e)={\cal N}(\bar{\bm{\mu}}_n,\bar{\bm{\sigma}}_n)\}$ are assumed to be standard normal distributions. Then the KL divergence term ${\cal L}_{KL}$ can be computed as closed form\cite{deng2023confidence}. To ensure that the loss ${\cal L}_{int}$ is differentiable, the Monte Carlo sampling method is used to approximate it, which is given by
\begin{equation}
\setlength{\abovedisplayskip}{2pt plus 1pt minus 1pt}
\setlength{\belowdisplayskip}{2pt plus 1pt minus 1pt}
{\cal L}_{bayes} \approx \frac{1}{K} \mathop\sum\limits_{k=1}^{K} {\cal L}_{int}(\bar{\cal {D}}^{s,e};\bm{r}_k^s, \bm{n}_k^e) + {\cal L}_{KL},
\label{eqn:monteSample}
\end{equation} 
where ${\bm r}_k^s={\bm \mu}_r^{s}+{\bm \sigma_r^{s}}\odot {\bm \epsilon}_k^{s}$, ${\bm n}_k^e={\bm \mu}_n^{e}+{\bm \sigma_n^{e}}\odot {\bm \epsilon}_k^{e}$, ${\bm \epsilon}_k^s$ and ${\bm \epsilon}_k^e$ are the $k$-th sample drawn from standard normal distributions. In this paper, ${\cal N}(\bm 0, \bm 1)$ and ${\cal N}(\bm 0, \bm {0.001})$ are empirically selected as the priors for LHUC and HUB parameters respectively. The location of speaker and environment transforms is empirically selected and fixed at the convolution subsampling module.
During adaptation, only one sample is drawn in Eqn.~(\ref{eqn:monteSample}). The Bayesian predictive inference integral is efficiently approximated by the expectation of the posterior distribution as $p(\tilde{\bm{Y}}^{s,e}|\tilde{\bm{X}}^{s,e},{\bm \mu}_{r}^{s},{\bm \mu}_{n}^{e})$ during recognition.

\begin{figure}
    \centering
    \includegraphics[scale=0.45]{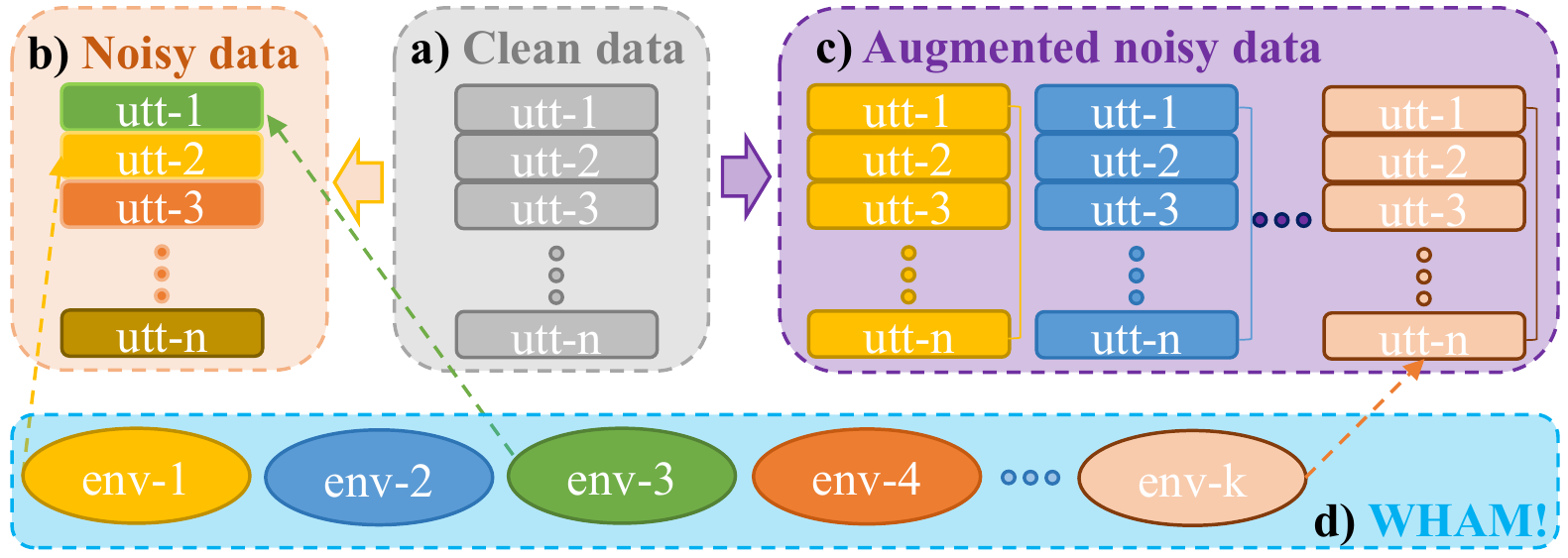}
    \setlength{\abovecaptionskip}{0pt plus 3pt minus 0pt}
    \caption{Noisy data simulation: a) Original Switchboard clean data, b) Non-augmented noise simulation, and c) Augmented noise simulation of Sec.~\ref{sec.expsetup}. d) Noise types in WHAM.}
    \label{fig:dataset}
    \vspace{-0.6cm}
\end{figure}

\begin{table*}[htbp]
\centering
\caption{Performance (WER\%) of adapted Conformer systems  with/without Bayesian learning evaluated on the noise corrupted, non-augmented Hub5'00, RT02 and RT03 sets, before and after external Transformer plus LSTM LM rescoring. “CHE”, “SWBD”, “FSH” and “O.V.” stand for “CallHome”, “Switchboard”, “Fisher” and “Overall” respectively. $\dagger$ and $\ast$ denote a statistically significant (MAPSSWE, $\alpha$=0.05) WER difference\cite{Gillick1989SomeSI} obtained over the baseline (sys. 1, 20) and the speaker adapted systems (sys. 3, 16, 21) respectively. The SNR and noise type combinations that appear in the train set are defined as "Seen" data, otherwise "Unseen" data.}
\vspace{-0.2cm}
\label{tab:table1_nonaugmented}
\resizebox{1.78\columnwidth}{!}{
\begin{tabular}{c|c|cc|c|c|ccc|cccc|ccc|ccc} %
\hline\hline
\multirow{2}{*}{ID} & 
\multirow{2}{*}{Method} & 
\multicolumn{2}{c|}{Adaptation} &
\multirow{2}{*}{\tabincell{c}{Adapt.\\Param.}} &
\multirow{2}{*}{\tabincell{c}{Language\\Model}} &   
\multicolumn{3}{c|}{Hub5'00 } & 
\multicolumn{4}{c|}{RT02} & 
\multicolumn{3}{c|}{RT03} & 
\multicolumn{3}{c}{ALL}
\\   \cline{7-19}
& & Speaker & Env. & & & CHE & SWBD & O.V. & SWBD1 & SWBD2 & SWBD3 & O.V. & FSH & SWBD & O.V. & Seen & Unseen & O.V. \\ \hline\hline
1 & Baseline &\xmark & \xmark  & - & \multirow{16}{*}{\xmark}& $36.4$ & $24.8$ & $30.6$ & $28.2$ & $34.8$ & $38.7$ & $34.3$ & $33.3$ & $38.8$ & $36.1$ & $22.9$ & $37.3$ & $34.2$ \\ \cline{1-5} \cline{7-19}
2 & \multirow{6}{*}{\tabincell{c}{Single\\Transform}} & LHUC & \xmark & \multirow{15}{*}{\tabincell{c}{Deter-\\ministic}} & & $35.0^{\dagger}$ & $24.2^{\dagger}$ & $29.6^{\dagger}$ & $27.1^{\dagger}$ & $33.8^{\dagger}$ & $37.9^{\dagger}$ & $33.3^{\dagger}$ & $31.9^{\dagger}$ & $38.5$ & $35.3^{\dagger}$ & $22.5^{\dagger}$ & $36.3^{\dagger}$ & $33.3^{\dagger}$\\ 
3 &  & HUB & \xmark  & & & $35.2^{\dagger}$ & $23.8^{\dagger}$ & $29.5^{\dagger}$ & $27.2^{\dagger}$ & $33.8^{\dagger}$ & $37.7^{\dagger}$ & $33.3^{\dagger}$ & $32.2^{\dagger}$ & $38.2^{\dagger}$ & $35.3^{\dagger}$ & $22.4^{\dagger}$ & $36.3^{\dagger}$ & $33.3^{\dagger}$ \\ 
4 &  & \xmark & LHUC & & & $35.9^{\dagger}$ & $23.8^{\dagger}$ & $29.9^{\dagger}$ & $27.2^{\dagger}$ & $34.2^{\dagger}$ & 38.5 & $33.7^{\dagger}$ & $32.5^{\dagger}$ & $38.1^{\dagger}$ & $35.4^{\dagger}$ & $22.3^{\dagger}$ & $36.6^{\dagger}$ & $33.5^{\dagger}$ \\ 
5 &  & \xmark & HUB & & & $36.1$ & $24.0^{\dagger}$ & $30.1^{\dagger}$ & $27.6^{\dagger}$ & $34.2^{\dagger}$ & 38.4 & $33.8^{\dagger}$ & $32.9^{\dagger}$ & $38.1^{\dagger}$ & $35.6^{\dagger}$ & $22.4^{\dagger}$ & $36.8^{\dagger}$ & $33.7^{\dagger}$ \\ 
6 &  & \multicolumn{2}{c|}{Joint LHUC} & & & $36.3$ & $24.6$ & $30.5$ & $28.1$ & $34.6$ & $38.7$ & $34.2$ & $33.3$ & $39.7$ & $36.6$ & $23.0$ & $37.5$ & $34.4$ \\
7 &  & \multicolumn{2}{c|}{Joint HUB} & & & $36.3$ & $24.7$ & $30.5$ & $28.1$ & $34.9$ & $39.0$ & $34.4$ & $33.1$ & $39.7$ & $36.5$ & $23.1$ & $37.5$ & $34.4$ \\ \cline{1-4} \cline{7-19}

8 & \multirow{4}{*}{\tabincell{c}{Linear\\Factorised \\ Adaptation \\ (LFA) }} & \multicolumn{2}{c|}{LHUC ($\beta=0.3$)} & & & $34.6^{\dagger\ast}$ & $23.2^{\dagger\ast}$ & $28.9^{\dagger\ast}$ & $\textbf{26.2}^{\dagger\ast}$ & $33.5^{\dagger}$ & $36.8^{\dagger\ast}$ & $32.5^{\dagger\ast}$ & $31.9^{\dagger}$ & $37.6^{\dagger\ast}$ & $34.9^{\dagger\ast}$ & $21.7^{\dagger\ast}$ & $35.6^{\dagger\ast}$ & $32.6^{\dagger\ast}$ \\

9 &  & \multicolumn{2}{c|}{LHUC ($\beta=0.5$)}  & & & $34.5^{\dagger\ast}$ & $23.0^{\dagger\ast}$ & $28.8^{\dagger\ast}$ & $26.3^{\dagger\ast}$ & $33.2^{\dagger\ast}$ & $36.9^{\dagger\ast}$ & $32.5^{\dagger\ast}$ & $31.3^{\dagger\ast}$ & $37.3^{\dagger\ast}$ & $34.4^{\dagger\ast}$ & $21.8^{\dagger\ast}$ & $35.4^{\dagger\ast}$ & $32.5^{\dagger\ast}$ \\ 

10 &  & \multicolumn{2}{c|}{LHUC ($\beta=0.7$)}  & & & $34.4^{\dagger\ast}$ & $\textbf{22.8}^{\dagger\ast}$ & $\textbf{28.6}^{\dagger\ast}$ & $\textbf{26.2}^{\dagger\ast}$ & $\textbf{33.1}^{\dagger\ast}$ & $\textbf{36.5}^{\dagger\ast}$ & $\textbf{32.3}^{\dagger}$ & $\textbf{31.2}^{\dagger\ast}$ & $\textbf{37.2}^{\dagger\ast}$ & $\textbf{34.3}^{\dagger\ast}$ & $\textbf{21.4}^{\dagger\ast}$ & $\textbf{35.3}^{\dagger\ast}$ & $\textbf{32.3}^{\dagger\ast}$ \\ 

11 &  & \multicolumn{2}{c|}{LHUC ($\beta=0.9$)} & & & $\textbf{34.3}^{\dagger\ast}$ & $23.1^{\dagger\ast}$ & $28.7^{\dagger\ast}$ & $26.9^{\dagger}$ & $\textbf{33.1}^{\dagger\ast}$ & $36.8^{\dagger\ast}$ & $32.6^{\dagger\ast}$ & $\textbf{31.2}^{\dagger\ast}$ & $37.6^{\dagger\ast}$ & $34.5^{\dagger\ast}$ & $21.7^{\dagger\ast}$ & $35.5^{\dagger\ast}$ & $32.5^{\dagger\ast}$ \\ \cline{1-4} \cline{7-19} 

12 & \multirow{4}{*}{\tabincell{c}{Cascaded\\Factorised\\Adaptation \\ (CFA)}} & LHUC & LHUC & & & $34.5^{\dagger\ast}$ & $23.6^{\dagger}$ & $29.1^{\dagger\ast}$ & $\textbf{26.5}^{\dagger\ast}$ & $33.3^{\dagger\ast}$ & $37.3^{\dagger\ast}$ & $32.8^{\dagger\ast}$ & $31.4^{\dagger\ast}$ & $37.6^{\dagger\ast}$ & $34.6^{\dagger\ast}$ & $21.7^{\dagger\ast}$ & $35.7^{\dagger\ast}$ & $32.7^{\dagger\ast}$ \\ 

13 &  & HUB & HUB & & & $\textbf{33.7}^{\dagger\ast}$ & $\textbf{23.0}^{\dagger\ast}$ & $\textbf{28.4}^{\dagger\ast}$ & $26.7^{\dagger\ast}$ & $\textbf{33.0}^{\dagger\ast}$ & $\textbf{36.2}^{\dagger\ast}$ & $\textbf{32.3}^{\dagger\ast}$ & $31.1^{\dagger\ast}$ & $\textbf{36.7}^{\dagger\ast}$ & $\textbf{34.0}^{\dagger\ast}$ & $\textbf{21.2}^{\dagger\ast}$ & $\textbf{35.1}^{\dagger\ast}$ & $\textbf{32.1}^{\dagger\ast}$ \\ 

14 &  & LHUC & HUB & & & $34.3^{\dagger\ast}$ & $23.1^{\dagger\ast}$ & $28.7^{\dagger\ast}$ & $26.8^{\dagger\ast}$ & $33.4^{\dagger\ast}$ & $37.0^{\dagger\ast}$ & $32.7^{\dagger\ast}$ & $\textbf{30.9}^{\dagger\ast}$ & $37.3^{\dagger\ast}$ & $34.2^{\dagger\ast}$ & $\textbf{21.2}^{\dagger\ast}$ & $35.5^{\dagger\ast}$ & $32.4^{\dagger\ast}$ \\ 

15 &  & HUB & LHUC & & & $33.8^{\dagger\ast}$ & $23.2^{\dagger\ast}$ & $28.5^{\dagger\ast}$ & $26.6^{\dagger\ast}$ & $33.6^{\dagger}$ & $37.0^{\dagger\ast}$ & $32.8^{\dagger\ast}$ & $31.2^{\dagger\ast}$ & $37.5^{\dagger\ast}$ & $34.4^{\dagger\ast}$ & $21.5^{\dagger\ast}$ & $35.5^{\dagger\ast}$ & $32.5^{\dagger\ast}$ \\ \hline

16 & \multirow{2}{*}{Single} & HUB & \xmark &\multirow{4}{*}{Bayesian} & \multirow{4}{*}{\xmark} & $34.4^{\dagger}$ & $23.1^{\dagger}$ & $28.8^{\dagger}$ & $27.0^{\dagger}$ & $33.3^{\dagger}$ & $37.1^{\dagger}$ & $32.8^{\dagger}$ & $31.3^{\dagger}$ & $37.6^{\dagger}$ & $34.6^{\dagger}$ & $21.8^{\dagger}$ & $35.6^{\dagger}$ & $32.6^{\dagger}$\\ 
17 &  & \multicolumn{2}{c|}{Joint LHUC} & & & $35.0^{\dagger}$ & $24.3^{\dagger}$ & $29.7^{\dagger}$ & 28.1 & $33.7^{\dagger}$ & $38.1^{\dagger}$ & $33.6^{\dagger}$ & $32.2^{\dagger}$ & $38.3^{\dagger}$ & $35.4^{\dagger}$ & $22.2^{\dagger}$ & $36.5^{\dagger}$ & $33.4^{\dagger}$  \\ \cline{2-2}
18 & LFA &\multicolumn{2}{c|}{LHUC ($\beta=0.7$)} & & & $33.1^{\dagger\ast}$ & $\textbf{22.4}^{\dagger\ast}$ & $\textbf{27.8}^{\dagger\ast}$ & $\textbf{25.9}^{\dagger\ast}$ & $32.8^{\dagger\ast}$ & $36.1^{\dagger\ast}$ & $32.0^{\dagger\ast}$ & $\textbf{30.4}^{\dagger\ast}$ & $36.9^{\dagger\ast}$ & $33.8^{\dagger\ast}$ & $20.9^{\dagger\ast}$ & $34.8^{\dagger\ast}$ & $31.8^{\dagger\ast}$ \\ \cline{2-2}
19 & CFA & HUB & HUB & & & $\textbf{33.0}^{\dagger\ast}$ & $22.6^{\dagger\ast}$ & $\textbf{27.8}^{\dagger\ast}$ & $26.1^{\dagger\ast}$ & $\textbf{32.0}^{\dagger\ast}$ & $\textbf{35.7}^{\dagger\ast}$ & $\textbf{31.6}^{\dagger\ast}$ & $30.6^{\dagger\ast}$ & $\textbf{36.6}^{\dagger\ast}$ & $\textbf{33.7}^{\dagger\ast}$ & $\textbf{20.7}^{\dagger\ast}$ & $\textbf{34.6}^{\dagger\ast}$ & $\textbf{31.6}^{\dagger\ast}$ \\ \hline

20 & Baseline &\xmark & \xmark & - & \multirow{5}{*}{\checkmark}& 35.9 & 23.8 & 29.9 & 27.3 & 33.9 & 37.3 & 33.2 & 32.5 & 38.0 & 35.3 & 21.9 & 36.4 & 33.3 \\ \cline{1-5} 

21 & \multirow{2}{*}{Single} & HUB & \xmark &\multirow{4}{*}{Bayesian} & & $34.0^{\dagger}$ & $22.4^{\dagger}$ & $28.2^{\dagger}$ & $26.2^{\dagger}$ & $32.5^{\dagger}$ & $35.9^{\dagger}$ & $31.9^{\dagger}$ & $30.3^{\dagger}$ & $36.8^{\dagger}$ & $33.7^{\dagger}$ & $20.7^{\dagger}$ & $34.8^{\dagger}$ & $31.8^{\dagger}$ \\ 
22 &  & \multicolumn{2}{c|}{Joint LHUC} & & & $34.5^{\dagger}$ & 23.5 & $29.0^{\dagger}$ & $26.8^{\dagger}$ & $33.0^{\dagger}$ & 37.2 & $32.7^{\dagger}$ & $31.5^{\dagger}$ & $37.5^{\dagger}$ & $34.6^{\dagger}$ & $20.9^{\dagger}$ & $35.8^{\dagger}$ & $32.7^{\dagger}$ \\ \cline{2-2}
23 & LFA &\multicolumn{2}{c|}{LHUC ($\beta=0.7$)} & & & $\textbf{31.6}^{\dagger\ast}$ & $21.8^{\dagger\ast}$ & $\textbf{26.8}^{\dagger\ast}$ & $25.4^{\dagger\ast}$ & $31.6^{\dagger\ast}$ & $34.8^{\dagger\ast}$ & $30.9^{\dagger\ast}$ & $29.7^{\dagger\ast}$ & $\textbf{35.3}^{\dagger\ast}$ & $32.6^{\dagger\ast}$ & $19.7^{\dagger\ast}$ & $33.6^{\dagger\ast}$ & $30.6^{\dagger\ast}$ \\ \cline{2-2}
24 & CFA & HUB & HUB & & & $32.0^{\dagger\ast}$ & $\textbf{21.5}^{\dagger\ast}$ & $\textbf{26.8}^{\dagger\ast}$ & $\textbf{24.9}^{\dagger\ast}$ & $\textbf{31.5}^{\dagger\ast}$ & $\textbf{34.1}^{\dagger\ast}$ & $\textbf{30.5}^{\dagger\ast}$ & $\textbf{29.4}^{\dagger\ast}$ & $\textbf{35.3}^{\dagger\ast}$ & $\textbf{32.4}^{\dagger\ast}$ & $\textbf{19.4}^{\dagger\ast}$ & $\textbf{33.3}^{\dagger\ast}$ & $\textbf{30.3}^{\dagger\ast}$ \\

\hline\hline
\end{tabular} 
}
\vspace{-0.6cm}
\end{table*}

\section{Experiments}
\subsection{Experimental Setup}
\label{sec.expsetup}
The widely used 300-hr Switchboard-1 conversational telephone corpus (LDC97S62)\cite{godfrey1992switchboard} containing 4804 speakers is utilized for training. The NIST 3.8-hr Hub5'00 (LDC2002S09, LDC2002T43), 6.4-hr RT02 (LDC2004S11), and 6.2-hr RT03 (LDC2007S10) test sets containing 80, 120, and 144 speakers respectively are adopted for performance evaluation. The publicly available noise WHAM database \cite{Wichern2019WHAM} which is recorded in non-stationary ambient environments such as restaurants, coffee shops, bars, parks, and office buildings is used as the noise source. Two protocols to simulate noise corrupted data are: {\bf 1) Non-augmented noise simulation} whereby each utterance is randomly exposed to one of multiple environments with a uniform distribution, as is shown in Fig.~\ref{fig:dataset}(b). {\bf 2) Augmented noise simulation} whereby each utterance is exposed to all different environments independently, as is shown in Fig.~\ref{fig:dataset}(c). For the noise corrupted training data, the 300-hr Switchboard-1 data is mixed with ten types of noise at signal-to-noise ratios (SNR) uniformly sampled from $\{-5, 0, 5, 10, 20\}$dB by the non-augmented simulation. For the noise corrupted evaluation sets, the test data is mixed with ten types of noise at SNRs uniformly sampled from $\{-15, -10, -5, 0, 5, 10, 20\}$dB. Three additive noise types used for the evaluation sets are also used in the training data simulation. Non-augmented noise simulation is applied to all three NIST Hub5'00, RT02 and RT03 sets for the first experiment presented in Table~\ref{tab:table1_nonaugmented}. 
To further analyse the improvements from Bayesian factorised adaptation, the experiments in Table~\ref{tab:table2_augmented} used 38-hr noise corrupted data derived by applying augmented noise simulation to the Hub5'00 set.

The ESPnet recipe \cite{guo2021recent} configured Conformer model comprised 12 encoder and 6 decoder layers, each with 256-dim 4-head attention and 2048 feed-forward hidden nodes. 80-dim Mel-filter bank plus 3-dim pitch parameters were used as input features, with byte-pair-encoding (BPE) tokens of size 2000 serving as decoder outputs. Two 2-D convolutional layers with stride 2 were included in the convolution subsampling module. 
SpecAugment \cite{Park2019SpecAugmentAS} was used for Conformer training. The initial learning rate of the Noam optimizer was 5.0. The dropout rate was set to 0.1, and the recognition model was averaged over the last ten epochs. The log-linearly interpolated external Transformer and Bi-LSTM language models (LMs), which were trained on the Switchboard and Fisher transcripts using cross-utterance contexts\cite{Sun2021TransformerLM}, were used for LM rescoring.

\vspace{-0.2cm}
\subsection{Experimental Results and Analysis}
\vspace{-0.1cm}
\textbf{Performance of Bayesian factorised adaptation} evaluated on the \textbf{non-augmented} noise corrupted test sets are shown in Table \ref{tab:table1_nonaugmented}. Several trends can be observed. \textbf{a)} Both the proposed linear factorised adaptation (LFA) (sys.8-11) and cascaded factorised adaptation (CFA) (sys.12-15) consistently outperformed the  un-adapted baseline (sys.1) and the adapted baselines (sys.2-4) considering only speaker or environment variability across all three test sets. The best operating point for LFA is $\beta=0.7$ (sys.10), while the best adaptation configuration of CFA is the “HUB-HUB" combination (sys.13). \textbf{b)} When Bayesian learning was further used to model LHUC and HUB parameters uncertainty, additional WER reductions of up to 0.8\% absolute (Hub5'00, sys.18 vs. sys.10) were consistently obtained using Bayesian factorised adaptations (sys.18,19) over that of the comparable non-Bayesian adapted systems (sys.10,13). \textbf{c)} Consistent WER reductions were retained after external LM rescoring. Overall statistically significant WER
reductions of \textbf{3.1\%}, \textbf{2.7\%}, \textbf{2.9\%} absolute (\textbf{10.4\%}, \textbf{8.1\%}, and \textbf{8.2\%} relative) were obtained by the proposed Bayesian CFA (sys.24) over the baseline Conformer (sys.20) on the noise corrupted Hub5’00, RT02 and RT03 test sets respectively. \textbf{d)} Joint speaker-environment adaptation (sys.6,7) using a single transform performed less well due to the lack of factorization between speaker and environment, and fragmentation of adaptation data. 

\begin{table}[htbp]
\centering
\vspace{-0.3cm}
\caption{Performance (WER\%) of adapted Conformer systems evaluated on the 38-hr noise corrupted and augmented Hub5'00 sets. $\dagger$ and $\ast$ denote statistically significant WER differences \cite{Gillick1989SomeSI} (MAPSSWE, $\alpha$=0.05) over the baselines (sys. 1, 20) and joint speaker-environment adaptation (sys. 7, 17, 22).}
\vspace{-0.3cm}
\label{tab:table2_augmented}
\resizebox{1.0\columnwidth}{!}{
\begin{tabular}{c|c|cc|c|c|cc|ccc} %
\hline\hline
\multirow{2}{*}{ID} & 
\multirow{2}{*}{Method} & 
\multicolumn{2}{c|}{Adaptation} &
\multirow{2}{*}{\tabincell{c}{Adapt.\\Param.}} &
\multirow{2}{*}{LM} &   
\multicolumn{5}{c}{38-hr Augmented Hub5'00 } 
\\   \cline{7-11}
& & Speaker & Env. &  & & CHE & SWBD & Seen & Unseen & O.V. \\ \hline\hline
1 & Baseline &\xmark & \xmark & - & \multirow{16}{*}{\xmark}& $36.6$ & $24.1$ & $29.3$ & $30.7$ & $30.4$\\ \cline{1-5} \cline{7-11}
2 & \multirow{6}{*}{\tabincell{c}{Single\\Transform}} & LHUC & \xmark  & \multirow{15}{*}{\tabincell{c}{Deter-\\ministic}} & & $35.4^{\dagger}$ & $24.0$ & $28.8^{\dagger}$ & $30.0^{\dagger}$ & $29.7^{\dagger}$\\ 
3 &  & HUB & \xmark &  & & $35.4^{\dagger}$ & $23.4^{\dagger}$ & $28.4^{\dagger}$ & $29.7^{\dagger}$ & $29.4^{\dagger}$  \\ 
4 &  & \xmark & LHUC &  & & $36.4$ & $23.8^{\dagger}$ & $29.1^{\dagger}$ & $30.4^{\dagger}$ & $30.1^{\dagger}$  \\ 
5 &  & \xmark & HUB & & & $35.9^{\dagger}$ & $23.8^{\dagger}$ & $28.8^{\dagger}$ & $30.3^{\dagger}$ & $29.9^{\dagger}$ \\ 
6 &  & \multicolumn{2}{c|}{Joint LHUC} & & & $35.1^{\dagger}$ & $23.3^{\dagger}$ & $28.3^{\dagger}$ & $29.5^{\dagger}$ & $29.2^{\dagger}$\\
7 &  & \multicolumn{2}{c|}{Joint HUB} & & & $34.9^{\dagger}$ & $23.2^{\dagger}$ & $28.0^{\dagger}$ & $29.3^{\dagger}$ & $29.0^{\dagger}$   \\ \cline{1-4} \cline{7-11}
8 & \multirow{4}{*}{LFA} & \multicolumn{2}{c|}{LHUC ($\beta=0.3$)} & & & $34.3^{\dagger\ast}$ & $22.8^{\dagger\ast}$ & $27.7^{\dagger\ast}$ & $28.8^{\dagger\ast}$ & $28.6^{\dagger\ast}$ \\
9 &  & \multicolumn{2}{c|}{LHUC ($\beta=0.5$)} & & & $34.1^{\dagger\ast}$ & $22.4^{\dagger\ast}$ & $27.2^{\dagger\ast}$ & $28.6^{\dagger\ast}$ & $28.3^{\dagger\ast}$  \\ 
10 &  & \multicolumn{2}{c|}{LHUC ($\beta=0.7$)} & & & $34.1^{\dagger\ast}$ & $\textbf{22.1}^{\dagger\ast}$ & $\textbf{27.0}^{\dagger\ast}$ & $\textbf{28.4}^{\dagger\ast}$ & $\textbf{28.1}^{\dagger\ast}$ \\ 
11 &  & \multicolumn{2}{c|}{LHUC ($\beta=0.9$)} & & & $\textbf{34.0}^{\dagger\ast}$ & $22.4^{\dagger\ast}$ & $27.3^{\dagger\ast}$ & $28.5^{\dagger\ast}$ & $28.3^{\dagger\ast}$ \\ \cline{1-4} \cline{7-11} 

12 & \multirow{4}{*}{CFA} & LHUC & LHUC & & & $34.3^{\dagger\ast}$ & $23.0^{\dagger}$ & $27.6^{\dagger\ast}$ & $29.0^{\dagger\ast}$ & $28.7^{\dagger\ast}$ \\ 
13 &  & HUB & HUB & & & $\textbf{33.8}^{\dagger\ast}$ & $\textbf{22.5}^{\dagger\ast}$ & $\textbf{27.2}^{\dagger\ast}$ & $\textbf{28.4}^{\dagger\ast}$ & $\textbf{28.2}^{\dagger\ast}$  \\  
14 &  & LHUC & HUB & & & $34.0^{\dagger\ast}$ & $22.9^{\dagger\ast}$ & $27.5^{\dagger\ast}$ & $28.8^{\dagger\ast}$ & $28.5^{\dagger\ast}$ \\ 
15 &  & HUB & LHUC & & & $34.0^{\dagger\ast}$ & $23.1^{\dagger}$ & $27.6^{\dagger\ast}$ & $28.8^{\dagger\ast}$ & $28.6^{\dagger\ast}$ \\ \hline

16 & \multirow{2}{*}{Single} & HUB & \xmark &\multirow{4}{*}{Bayes} & \multirow{4}{*}{\xmark}& $34.9^{\dagger}$ & $23.2^{\dagger}$ & $28.1^{\dagger}$ & $29.3^{\dagger}$ & $29.0^{\dagger}$\\ 
17 &  & \multicolumn{2}{c|}{Joint HUB} & & & $34.5^{\dagger}$ & $22.8^{\dagger}$ & $27.6^{\dagger}$ & $28.9^{\dagger}$ & $28.6^{\dagger}$ \\ \cline{2-2}
18 & LFA & \multicolumn{2}{c|}{LHUC ($\beta=0.7$)} & & & $33.4^{\dagger\ast}$ & $\textbf{21.9}^{\dagger\ast}$ & $\textbf{26.6}^{\dagger\ast}$ & $\textbf{28.0}^{\dagger\ast}$ & $\textbf{27.7}^{\dagger\ast}$ \\ 
19 & CFA & HUB & HUB & & & $\textbf{33.3}^{\dagger\ast}$ & $22.3^{\dagger\ast}$ & $27.0^{\dagger\ast}$ & $\textbf{28.0}^{\dagger\ast}$ & $27.8^{\dagger\ast}$ \\  \hline

20 & Baseline &\xmark & \xmark & - & \multirow{5}{*}{\checkmark}& $36.0$ & $23.3$ & $28.5$ & $30.1$ & $29.7$   \\ \cline{1-5} 
21 & \multirow{2}{*}{Single} & HUB & \xmark &\multirow{4}{*}{Bayes} & & $34.4^{\dagger}$ & $22.4^{\dagger}$ & $27.6^{\dagger}$ & $28.5^{\dagger}$ & $28.4^{\dagger}$  \\ 
22 &  & \multicolumn{2}{c|}{Joint HUB}  & & & $33.7^{\dagger}$ & $21.8^{\dagger}$ & $26.5^{\dagger}$ & $28.1^{\dagger}$ & $27.7^{\dagger}$ \\ \cline{2-2}
23 & LFA & \multicolumn{2}{c|}{LHUC ($\beta=0.7$)} & & & $32.4^{\dagger\ast}$ & $\textbf{21.0}^{\dagger\ast}$ & $\textbf{25.6}^{\dagger\ast}$ & $27.0^{\dagger\ast}$ & $\textbf{26.7}^{\dagger\ast}$  \\
24 & CFA & HUB & HUB & & & $\textbf{32.3}^{\dagger\ast}$ & $21.1^{\dagger\ast}$ & $26.0^{\dagger\ast}$ & $\textbf{26.9}^{\dagger\ast}$ & $\textbf{26.7}^{\dagger\ast}$  \\ 
\hline\hline
\end{tabular}}
\vspace{-0.4cm}
\end{table}

\noindent
\textbf{Performance of Bayesian factorised adaptation} evaluated on the \textbf{augmented} noise corrupted Hub5'00 set (38-hr) are shown in Table~\ref{tab:table2_augmented}. The trends found in Table~\ref{tab:table1_nonaugmented} were still retained. Overall absolute WER reductions of \textbf{3.0\%} and \textbf{1.0\%} were obtained by the Bayesian CFA (sys.24) over the baseline (sys.20) and the joint speaker-environment adapted (sys.22) systems. 

\noindent
\textbf{Potential for rapid adaptation:}In the experiments of Table~\ref{tab:table3_analysis} where either the speaker transforms are estimated using speaker level data in mismatched environments (sys.5),  or the environment transforms are learned using environment level data with mismatched speakers (sys.6), or both being mismatched against the test data being adapted to (sys.7,8), factorised speaker-environment adaptation consistently produced absolute WER reductions of \textbf{1.1\%-2.5\%} over the baseline un-adapted Conformer (sys.1). In particular, the CFA factorised adaptation with both speaker and environment mismatches produced performance comparable to speaker only adaptation using the matched environment (sys.7 vs. sys.2). These results suggest that the proposed flexible factorization framework allows the separately acquired speaker and environment homogeneity by factorization to be exploited for rapid adaptation to unseen speaker-environment combinations.

\begin{table}[htbp]
\centering
\vspace{-0.2cm}
\caption{Performance (WER\%) of Bayesian factorised adaptation using matched or mismatched transforms evaluated on the 3.8-hr subset of 38-hr noise corrupted Hub5’00 set. Five mismatched conditions are randomly selected for each utterance.}
\vspace{-0.35cm}
\label{tab:table3_analysis}
\resizebox{1.0\columnwidth}{!}{
\begin{tabular}{c|c|c|c|ccc} %
\hline\hline
\multirow{2}{*}{ID} & 
\multirow{2}{*}{Method} & 
\multirow{2}{*}{Speaker Transform} & 
\multirow{2}{*}{Env. Transform} & 
\multicolumn{3}{c}{Hub5'00 ($mean {\pm std}$)} 
\\   \cline{5-7}
& & & & CHE & SWBD & O.V. \\ \hline\hline
1 & Baseline & - & - & $37.7$ & $24.2$ & $31.0$ \\ \hline

2 & \multirow{2}{*}{\tabincell{c}{HUB \\(Spk. adapt)}} & Matched Env. & - & $35.9$ & $23.4$ & $29.7$ \\ 
3 &  & Mismatched Env. & - & $37.0_{\pm 0.27}$ & $24.3_{\pm 0.31}$ & $30.7_{\pm 0.23}$ \\ \hline
4 & \multirow{4}{*}{\tabincell{c}{CFA\\(HUB-HUB)}} & Matched Env. & Matched Spk. & $34.0$ & $22.1$ & $28.1$ \\ 
5 & & Mismatched Env. & Matched Spk. & $34.4_{\pm0.22}$ & $22.4_{\pm0.25}$ & $28.5_{\pm 0.15}$ \\ 
6 & & Matched Env.& Mismatched Spk. & $34.9_{\pm0.20}$ & $22.8_{\pm0.15}$ & $28.9_{\pm 0.19}$ \\ 
7 & & Mismatched Env. & Mismatched Spk. & $35.8_{\pm0.23}$ & $23.4_{\pm0.26}$ & $29.6_{\pm 0.20}$  \\ \hline
8 & LFA ($\beta=0.7$) & Mismatched Env. & Mismatched Spk. & $36.3_{\pm0.26}$ & $23.6_{\pm0.32}$ & $29.9_{\pm 0.22}$  \\ 

\hline\hline
\end{tabular} 
}
\end{table}
\vspace{-0.5cm}
\section{Conclusions}
\vspace{-0.05cm}
The paper proposed a novel Bayesian factorised speaker-environment adaptive training and test time unsupervised adaptation approach for Conformer models. Compact transformations were used to model speaker and environment level characteristics separately, which were linearly or hierarchically combined to represent any seen or unseen speaker-environment combination. Bayesian learning was further utilized to model the adaptation parameter uncertainty. Experiments on the 300-hour WHAM noise corrupted Switchboard corpus showed that the proposed Bayesian factorised adaptation produced up to 3.1\% absolute (10.4\% relative) WER reductions over the un-adapted baseline Conformer system. 

\vspace{-0.2cm}
\section{Acknowledgements}
\vspace{-0.05cm}
This research is supported by Hong Kong RGC GRF grant No. 14200021, 14200220, Innovation \& Technology Fund grant No. ITS/254/19 and ITS/218/21, National Natural Science Foundation of China (NSFC) Grant 62106255, and Youth Innovation Promotion Association CAS Grant 2023119.

\bibliographystyle{IEEEtran}
\bibliography{mybib}

\end{document}